\documentclass[a4paper,11pt]{article}
\usepackage{geometry}\usepackage[utf8]{inputenc}
\usepackage{amsmath}\usepackage{amssymb}
\usepackage{xcolor}

\newcommand{\fsl}[1]{#1\kern-0.50em/}
\newcommand{\fsnabla}{\nabla\kern-0.75em/}
\newcommand{\di}{\text{d}}

\title{ Super-Yang-Mills Theory in SIM(1) Superspace }
\author{ Ji\v{r}\'{i} Voh\'{a}nka$^1$ and Mir Faizal$^2$ \\ 
\\ $^1$Masaryk University, Department of Theoretical Physics and Astrophysics\\
Kotl\'{a}\v{r}sk\'{a} 267/2, 611 37 Brno, Czech Republic \\ 
$^2$Department of Physics and Astronomy,   University of Waterloo, \\  Waterloo, 
Ontario N2L 3G1, Canada 
}
\date{}
\begin{document}

\maketitle

\begin{abstract}
In this paper, we will analyse three dimensional 
supersymmetric Yang-Mills theory coupled to matter fields in  $SIM(1)$ superspace formalism. 
The original  theory which is invariant under the full Lorentz group  has $\mathcal{N} =1$ supersymmetry. 
However, when we break the Lorentz symmetry down to $SIM(1)$ group,  
the $SIM(1)$ superspace will break half the supersymmetry of the original theory. 
Thus, the resultant theory in $SIM(1)$ superspace will have $\mathcal{N} =1/2$ supersymmetry. This is the
first time    that   $\mathcal{N} =1 $ supersymmetry will be broken down to $\mathcal{N} =1/2$ supersymmetry, for a 
three dimensional theory, 
on a manifold without a boundary. 
 This is because 
it is not possible to use non-anticommutativity 
 to break 
  $\mathcal{N} =1$ supersymmetry down to $\mathcal{N} =1/2$ supersymmetry in three dimensions. 
\end{abstract}

\section{Introduction}
  Lorentz symmetry is one of the most important symmetries in nature. However, there    are  strong theoretical 
indications it might only be an effective symmetry and it might break   at Planck scale.  
These theoretical 
indications come from  various approaches to quantum gravity. 
For example, in string theory 
the unstable  perturbative string vacuum is expected to break   Lorentz symmetry  \cite{1a}-\cite{2a}. 
This happens as in this case certain 
  tensors acquire non-zero vacuum expectation values, which in turn induce a 
preferential direction in spacetime. In fact, as  string theory is related to noncommutativity and noncommutativity is expected to break 
Lorentz symmetry, it is not a surprise that unstable  perturbative string vacuum can break Lorentz symmetry 
 \cite{01}-\cite{02}.
In most approaches to quantum gravity the Lorentz symmetry is expected to break at  Planck scale \cite{04}. 
One way to observe that is to note that the 
  gravity is not renormalizable. It can only be made  
renormalizable by adding higher terms to original action \cite{4}, which in turn break the  
unitarity of the   theory \cite{5}. 
The unitarity can be preserved by taking a 
different Lifshitz scaling for space and time, thus,   adding higher order spatial derivatives to the theory 
without adding any term containing higher order temporal derivative. 
This theory is called Ho\v{r}ava-Lifshitz gravity, and it obviously breaks Lorentz symmetry \cite{6}-\cite{7}.
In fact, even in loop quantum gravity, Lorentz symmetry is expected to break at Planck scale \cite{8}-\cite{9}. 
There have been attempts to study a model where a system is only invariant under subgroups
of the Lorentz group, such that this subgroup still preserves enough symmetry 
for the 
constancy of the velocity of light. 
This theory is called  very special relativity (VSR) \cite{vsr}.
In this theory 
 the whole Lorentz group is recovered if CP symmetry is also postulated as a symmetry of the system.
 Two subgroups of Lorentz group called the $SIM(2)$  and $HOM(2)$ have been studied in this 
regard. The advantage of using these subgroups is that  the dispersion
relations, time delay and all classical tests of special relativity  are valid for these subgroups. 

The VSR   can also be realized as the part of the Poincare 
symmetry preserved on a noncommutative Moyal plane with light-like noncommutativity \cite{a1}. 
In fact, the three subgroups relevant to the   VSR can also be realized in the noncommutative spacetime setting.
Quantum field theory with abelian gauge symmetry have been studied in spacetime with the symmetry group corresponding to  VSR 
\cite{a}. This work has been recently generalized to include non-abelian gauge theories \cite{b}. 
Four dimensional  supersymmetric theories have been analysed in $SIM(2)$ \cite{sup}. In fact, a superspace construction \cite{sups}
and supergraph rules \cite{supa} for such 
theories have also been developed. This $SIM(2)$ superspace formalism has  been  used for analysing gauge theories 
\cite{supg}.

It may be noted that if the 
  Lorentz invariant is broken down to invariance under $SIM(2)$ group, the resultant 
  $SIM(2)$ superspace breaks half the supersymmetry of the original theory.
  Thus, if we modify     a four dimensional   Lorentz invariant theory with   $\mathcal{N} =1 $ supersymmetry, 
  to $SIM(2)$ superspace, the resultant theory has $\mathcal{N} =1/2 $ supersymmetry.
  The terminology  $\mathcal{N} =1/2 $ supersymmetry is borrowed from non-anticommutative deformation of a theory in four dimensions. 
  This is because in four dimensions, it is also possible to 
  break half the supersymmetry of a theory by deforming the theory to  a non-anticommutative superspace \cite{non}-\cite{non1}.  
  So, if non-anticommutativity is imposed on a four dimensional theory with $\mathcal{N} =1 $ supersymmetry, the resultant theory 
  is called a theory with $\mathcal{N} =1/2 $ supersymmetry, as it preserves only half the supersymmetry of the original theory. 
The breaking of the Lorentz group down to $SIM(2)$ group also breaks half the supersymmetry of the original Lorentz invariant theory. 
So, the amount of  supersymmetric broken by breaking the Lorentz symmetry of a  theory to $SIM(2)$ superspace is the same as
  the amount of  supersymmetric broken by deforming it by imposing 
  non-anticommutativity. Thus, if the Lorentz symmetry of  four dimensional theory with $\mathcal{N} =1  $ supersymmetry
  is broken down to $SIM(2)$ group, the resultant theory will also be called called a theory with
  $\mathcal{N} =1/2 $ supersymmetry. 
  
  It is not possible  to break the supersymmetry of a three dimensional theory from 
  $\mathcal{N} =1 $ supersymmetry to $\mathcal{N} =1/2 $ supersymmetry by deforming it to a non-anticommutative superspace. 
  This is because there are not enough anticommutative degrees of freedom to perform such a deformation. Any non-anticommutative  deformation 
  of a three dimensional supersymmetric theory with $\mathcal{N} =1  $ supersymmetry will break all the supersymmetry 
  of the theory. It is possible to break the supersymmetry of a three dimensional theory with $\mathcal{N} = 2 $ supersymmetry 
  down to $\mathcal{N} = 1$ supersymmetry by imposing non-anticommuativity \cite{nonnon}.
However,  a three dimensional theory with $\mathcal{N} =1/2$ supersymmetry can be constructed on a manifold with a boundary \cite{b1}-\cite{b2}. 
  This is because the boundary effects break half the supersymmetry of the original theory. So, if a theory 
 has  $\mathcal{N} = 1$ supersymmetry  in absence of a boundary , the same  theory  will only have 
  $\mathcal{N} =1/2$ supersymmetry in presence of a boundary. Furthermore, in presence of a boundary, we
   can also use projections   to construct a theory with  $\mathcal{N} = (1, 1) $ supersymmetry. 
As both the   boundary effects and non-anticommutativity  breaks half the supersymmetry, 
 it is possible to use a  different projection to impose non-anticommutativity from the projection used to preserve half  
 the supersymmetry on the boundary.
So,   for a three dimensional theory with 
 $\mathcal{N} = (1, 1) $ supersymmetry, it  is  also possible break  the supersymmetry  down to $\mathcal{N} =(1/2, 0)  $
 supersymmetry by combining non-anticommutativity with boundary effects  \cite{bou}. 
  However, the advantage of using $SIM(1)$ superspace is that, we will be able  construct a three dimensional theory with
  $\mathcal{N} =1/2 $ supersymmetry by modifying  
  a theory with $\mathcal{N} =1 $ supersymmetry  on a manifold without a boundary. Thus, it is the first time a three dimensional theory 
  with $\mathcal{N} =1/2 $ supersymmetry will be constructed from a theory  $\mathcal{N} =1 $ supersymmetry on a manifold without a boundary.
  
\section{Subgroup of the Lorentz group preserving light-like direction}

A Very Special Relativity \cite{vsr} works with space-time symmetry reduced to a subgroup of the Lorentz group.
In four dimensions the largest such subgroup is $SIM(2)$ a group that preserves fixed light-like vector up to its rescaling.
It is possible to consider subgroups of the Lorentz group determined by such condition also in dimensions other than four.
We will examine this possibility in this section.
 
An infinitesimal transformation of a vector $x$ under the group $SO(D-1,1)$ is given as
$
 \delta x^a = \omega^a{}_b x^b, 
$
where $\omega^a{}_b$ are infinitesimal parameters chosen such that the size of the vector is not changed. 
This means that $0 = \delta(x^2) = x^a(\omega^c{}_a\eta_{cb}+\omega^c{}_b\eta_{ac})x^b$, because this must hold 
for any vector, the expression inside brackets must vanish. If we use the metric $\eta_{ab}$ for rising and lowering of
indices we get the condition that $\omega$ must be antisymmetric $\omega^{ab}+\omega^{ba} = 0$.

In addition to invariance of size of vectors we impose the condition that some null vector $n$ is preserved 
up to rescaling. This can be written as 
\begin{equation}\label{sg:Dc}
 \delta n^a = \omega^{ab}n_b = -2A n^a, \qquad A\in\mathbb{R}.
\end{equation}
It is convenient to work with light-cone coordinates $x^{\pm}=\frac{1}{\sqrt{2}}(x^0\pm x^1)$, where the
indices take values in the set $+,-,2\ldots,D-1$ and metric is
\begin{equation}
 \eta = \begin{pmatrix}&-1&\\-1&&\\&&\mathbf{1}_{D-2}\end{pmatrix},
\end{equation}
where $\mathbf{1}_{D-2}$ denotes $(D-2)\times(D-2)$ unit matrix. 
We choose the null-vector such that it has only one nonzero coordinate $n^+=-n_-=\frac{1}{\sqrt{2}}$, and the remaining coordinates vanish
$n^-=-n_+=0$, $n^a=n_a=0$ for $a=2,\ldots,D-1$.
The condition Eq.  \eqref{sg:Dc} then leads to
\begin{align}
 \omega^{+b}n_b &= A n^+ &&\Rightarrow & \omega^{+-} &= 2A, 
\nonumber\\
 \omega^{-b}n_b &= A n^- &&\Rightarrow & 0 &= 0, 
\nonumber\\
 \omega^{ab}n_b &= A n^a &&\Rightarrow & \omega^{a-} &= 0, & \text{for }a=2,\ldots,D-1.
\end{align}
The third condition is the only one that restricts the infinitesimal parameters, it sets $D-2$ of them to zero.
Thus the dimension of the resulting group is $\dfrac{D(D-1)}{2}-(D-2)$.

In the case of $D=3$ the dimension of the group is $2$ and the matrix $\omega$ has the form
\begin{align}
\omega^{ab} &= 
 \begin{pmatrix}0&\omega^{+-}&\omega^{+2}\\\omega^{-+}&0&\omega^{-2}\\\omega^{2+}&\omega^{2-}&0\end{pmatrix} =
 \begin{pmatrix}0&2A&-\sqrt{2}B\\-2A&0&0\\\sqrt{2}B&0&0\end{pmatrix}
\end{align}
where $A,B\in\mathbb{R}$. 
The exponentiation of the infinitesimal transformation gives the transformation
\begin{align}\label{sg:3}
 x^+ &\rightarrow e^{-2A} x^+ - \sqrt{2} e^{-A} B x^2 + B^2 x^-,
&
 x^- &\rightarrow e^{2A} x^-,
&
 x^2 &\rightarrow x^2 - \sqrt{2} e^{A} B x^-.
\end{align}

Another way how to arrive to this group is to represent vectors by two-dimensional symmetric matrices
\begin{equation}
 x = \begin{pmatrix}x^0+x^1&x^2\\x^2&x^0-x^1\end{pmatrix} 
 = \begin{pmatrix}\sqrt{2}x^+&x^2\\x^2&\sqrt{2}x^-\end{pmatrix},
\end{equation}
the size of the vector is $x^2=-\det x$ and the size-preserving transformations are given as
\begin{equation}
 x' = gxg^T, \qquad g\in SL(2,\mathbb{R}).
\end{equation}

Any null-vector can be written as $n^{\alpha\beta}=\xi^{\alpha}\xi^{\beta}$ where the commuting spinor $\xi$ is 
determined uniquely up to a sign. A convenient choice of a null vector is to choose $\xi^+=1$, $\xi^-=0$.
The condition that this null-vector is preserved up to a rescaling can be now written as
\begin{equation}
 g\xi = \pm e^{-A}\xi, \qquad A\in\mathbb{R}.
\end{equation}
Only matrices from $SL(2,\mathbb{R})$ that satisfy this criteria are 
\begin{equation}\label{sg:3tm}
 g = \pm\begin{pmatrix}e^{-A}&-B\\0&e^{A}\end{pmatrix}, 
\end{equation}
the meaning of $A$ and $B$ is the same as in Eq. \eqref{sg:3}. 

The difference between $D=3$ and $D=4$ is that in the case of $D=4$ we worked with complex matrices from 
$SL(2,\mathbb{C})$, while in the case of $D=3$ we have real matrices from $SL(2,\mathbb{R})$. 
In the four dimensional case we called this group $SIM(2)$ a group of similarity transformations in two dimensions 
(consisting of rotation, scaling and shift). 
In our $D=3$ case we can identify this group with a group of orientation preserving similarity
transformations in one dimension (consisting of scaling and shift). In order to show that we identify a point in
one-dimensional space determined by coordinate $z$ with a point in projective space $\mathbb{R}\textrm{P}^1$ represented
by $\left(\begin{matrix}z\\1\end{matrix}\right)$. An action of the group given by left multiplication by $g$
then gives
\begin{equation}
 \begin{pmatrix}z\\1\end{pmatrix} 
 \quad\rightarrow\quad 
 g\begin{pmatrix}z\\1\end{pmatrix}
 =\pm\begin{pmatrix}e^{-A}z-B\\e^{A}\end{pmatrix}
 \sim\begin{pmatrix}e^{-2A}z-e^{-A}B\\1\end{pmatrix}
 =\begin{pmatrix}z'\\1\end{pmatrix}.
\end{equation}
The change $z\rightarrow z'$ indeed describes the orientation preserving similarity transformation.

\section{Three dimensional Supersymmetry} \label{susy3}
 
In this section, we will study three dimensional superspace. 
In the  Lorentz invariant theory,  $\mathcal{N} =1$ supersymmetry will be generated by  
\begin{equation}
  Q_\alpha  = \partial_\alpha - (\gamma^a\theta)_\alpha \partial_a
  = \partial_\alpha + \gamma^a_{\alpha\beta}\theta^\beta \partial_a.
\end{equation}
This generator of $\mathcal{N} =1$ supersymmetry in three dimensions commutes with the super-derivative $D_\alpha$, where 
\begin{equation}
 D_\alpha = \partial_\alpha + (\gamma^a\theta)_\alpha \partial_a
  = \partial_\alpha - \gamma^a_{\alpha\beta}\theta^\beta \partial_a. 
\end{equation}
The full supersymmetry algebra that $Q_\alpha$ and $D_\alpha$ satisfy is given by 
\begin{align}
 \{Q_\alpha,Q_\beta\} &= 2\gamma^a_{\alpha\beta}\partial_a,
&
 \{Q_\alpha,D_\beta\} &= 0,
&
 \{D_\alpha,D_\beta\} &= -2\gamma^a_{\alpha\beta}\partial_a. 
\end{align}

Now as was shown in Eq.  \eqref{sg:3tm}, the $SIM(1)$-transformation of spinors is given as
\begin{align}
 \begin{pmatrix} \psi'^+ \\ \psi'^- \end{pmatrix} &= 
 \begin{pmatrix} e^{-A}&-B \\ 0&e^{A} \end{pmatrix}
 \begin{pmatrix} \psi^+ \\ \psi^- \end{pmatrix}
&&\Leftrightarrow&
 \begin{pmatrix} \psi'_+ \\ \psi'_- \end{pmatrix} &= 
 \begin{pmatrix} e^{A}&0 \\ B&e^{-A} \end{pmatrix}
 \begin{pmatrix} \psi_+ \\ \psi_- \end{pmatrix}
\end{align}
with $A,B\in\mathbb{R}$.
Spinors that satisfy the condition
\begin{equation}
 \fsl{n}\psi = 0 \qquad \Rightarrow \qquad \psi = \begin{pmatrix} 0 \\ \psi_- \end{pmatrix}
\end{equation}
consist a space that is invariant under $SIM(1)$ transformations. 
Let us denote the space of all spinors as $\mathcal{S}$, and the invariant space that we have just described as 
$\mathcal{S}_{\text{invariant}}$. 
We also define a space $\mathcal{S}_{\text{quotient}} = \mathcal{S}/\mathcal{S}_{\text{invariant}}$.
A convenient description of this space is provided by choosing a representative 
\begin{equation}
 \psi = \begin{pmatrix} \psi_+ \\ 0 \end{pmatrix}.
\end{equation}
in each equivalence class.
Both spaces $\mathcal{S}_{\text{invariant}}$ and $\mathcal{S}_{\text{quotient}}$ carry a representation of the
$SIM(1)$-group, they transform as
\begin{align}
 \begin{pmatrix} 0 \\ \psi'_- \end{pmatrix} &= e^{-A} \begin{pmatrix} 0 \\ \psi'_- \end{pmatrix},
&
 \begin{pmatrix} \psi'_+ \\ 0 \end{pmatrix} &= e^{A} \begin{pmatrix} \psi'_+ \\ 0 \end{pmatrix}.
\end{align}
Thus we have two distinct one-dimensional representations.

The reduction of supersymmetry will be done following the same steps as in the four dimensional case \cite{sup}, \cite{sups}.
We can summarize the necessary steps as:
\begin{itemize}
\item The space-time symmetry will be reduced to $SIM(1)$-subgroup.
\item The supersymmetry transformations will be reduced to those that correspond to symmetry generator 
$\bar{\epsilon}Q$ with anticommuting parameter $\epsilon$ satisfying the condition $\fsl{n}\epsilon=0$.
There will remain only one supersymmetry generator proportional to $\fsl{n}Q$. 
It will transforms under $SIM(1)$ in the same way as spinors from $\mathcal{S}_{\text{quotient}}$.
\item Only the anticommuting $\theta$ coordinates that satisfy $\fsl{n}\theta=0$ are kept in the superspace.
There will be only one anticommuting coordinate. 
It will transform under $SIM(1)$ in the same way as spinors from $\mathcal{S}_{\text{invariant}}$.
\item The covariant derivatives are reduced in a similar way as supersymmetry generators.
There are two things we have to take care of. Firstly, we only keep covariant spinor derivatives that are 
proportional to $\fsl{n}D$. Secondly, because the resulting superspace is reduced we have to make a projection that
removes anticommuting coordinates that are no longer part of it (ie. projection that sets 
$\fsl{n}\theta=0$). There will be one spinor covariant derivative that will transform under $SIM(1)$ in the same way as 
spinors from $\mathcal{S}_{\text{quotient}}$.
\end{itemize}

The above steps can be easily done if we introduce another null-vector $\tilde{n}$ that satisfies the relation 
$n\cdot\tilde{n}=1$. This allows us to define projectors that split any spinor into two parts
\begin{equation}
\psi = \frac{1}{2}\fsl{\tilde{n}}\fsl{n}\psi + \frac{1}{2}\fsl{n}\fsl{\tilde{n}}\psi
\qquad\Leftrightarrow\qquad
\psi_\alpha = -\tilde{n}_{\alpha\beta}n^{\beta\gamma}\psi_\gamma -n_{\alpha\beta}\tilde{n}^{\beta\gamma}\psi_\gamma.
\end{equation}
With the choice of $n$ and $\tilde{n}$ in which only the components $n^{++}=i$, $\tilde{n}^{--}=-i$ are nonzero
we get
\begin{align}
 \frac{1}{2}\fsl{\tilde{n}}\fsl{n}\psi &= \begin{pmatrix}\psi_+\\0\end{pmatrix},
&
 \frac{1}{2}\fsl{n}\fsl{\tilde{n}}\psi &= \begin{pmatrix}0\\\psi_-\end{pmatrix}.
\end{align}
The supersymmetry generator $S$, the anticommuting superspace coordinate $\zeta$ and spinor derivative $d$ are
defined as
\begin{align}\label{nnproj}
 S &= \frac{1}{2}\fsl{\tilde{n}}\fsl{n}Q,
&
 \zeta &= \frac{1}{2}\fsl{n}\fsl{\tilde{n}}\theta,
&
 d &= \frac{1}{2}\fsl{\tilde{n}}\fsl{n}D\big\vert_{\fsl{n}\theta = 0}.
\end{align}
Each of them have only one nonzero component
\begin{align}
 S_+ &= \partial_+ + i\zeta_- \partial_{++}
&
 \zeta_- &= \theta_-,
&
 d_+ &= \partial_+ - i\zeta_- \partial_{++}
\end{align}
and they satisfy
\begin{align}
 \{S_+,S_+\} &= 2\partial_{++},
&
 \{S_+,d_+\} &= 0,
&
 \{d_+,d_+\} &= -2\partial_{++},
&
 \partial_+ \zeta_- &= -i.
\end{align}

It may be noted that this modification breaks half the supersymmetry of the original theory. 
Thus, as our original theory had $\mathcal{N} =1$ supersymmetry, the resultant theory after this modification only has 
 $\mathcal{N} =1/2$ supersymmetry. 
 Unlike the four dimensional case \cite{non}-\cite{non1}, 
 we cannot break the supersymmetry of a three dimensional theory with $\mathcal{N} =1$ supersymmetry
 to  $\mathcal{N} =1/2$ supersymmetry by using non-anticommutativity. This is because in four dimensions there are enough degrees of freedom to 
 partially break the  $\mathcal{N} =1$ supersymmetry.  For a four dimensional theory with 
  $\mathcal{N} =1 $ supersymmetry, there are four independent anticommutating coordinates. So,  if
  non-anticommutativity is imposed between  two of them, still the supersymmetry corresponding to the other two 
 is preserved. However, for  three dimensional theory with 
   $\mathcal{N} =1 $ supersymmetry
 there are only two independent anticommutating coordinates, and so, any  non-anticommutativity
 will break all the supersymmetry of such a  three dimensional theory. 
 Hence,   a three dimensional theory with 
 $\mathcal{N} =1/2$ supersymmetry can be obtained   by breaking the Lorentz symmetry down to  $SIM(1)$ group.

\section{Superfield decomposition}

In this section we are going to establish correspondence between superfields that appear in $SO(2,1)$ superspace
and superfields that we use in $SIM(1)$ superspace. There are two things we have to resolve in order to establish
this correspondence.
Firstly, the $SO(2,1)$ superspace is bigger than $SIM(1)$ superspace. This means that if we write a $SO(2,1)$ theory in
$SIM(1)$ superspace then to each $SO(2,1)$ superfield there will correspond multiple $SIM(1)$ superfields,
otherwise we lose some degrees of freedom. In fact, we will observe that for each $SO(2,1)$ superfield there are two
  $SIM(1)$ superfields. 
Secondly, if the $SO(2,1)$ superfield carries some space-time indices then we have have to handle them specially,
otherwise we will get $SIM(1)$ superfields that transform in a very complicated way under $SIM(1)$ group.

Let us start with a scalar $SO(2,1)$ superfield $\Phi$. The projections
\begin{align}
 \phi &= \Phi \vert_{\theta_+ = 0},
&
 \tilde{\phi}_- &= ( D_- \Phi ) \vert_{\theta_+ = 0},
\end{align}
contain all information carried by $\Phi$. This is most easily seen from the fact that the superfield $\Phi$ could be
written as 
\begin{equation}
 \Phi = \phi -i\theta_+\left(\tilde{\phi}_- + i \theta_- \partial_{+-}\phi \right).
\end{equation}
The $SIM(1)$ rotations change these superfields as (prime denotes transformed quantities)
\begin{align}\label{tsimt}
 \phi'(x',\theta') &= \phi(x,\theta),
&
 \tilde{\phi}'_-(x',\theta') &= e^{-A}\tilde{\phi}_-(x,\theta) + B \phi(x,\theta).
\end{align}
The superfield $\phi$ transforms nicely but the superfield $\tilde{\phi}_-$ transforms into combination of both
$\phi$ and $\tilde{\phi}_-$ which makes it unsuitable for description of $SIM(1)$ theories because it makes 
$SIM(1)$ invariance non-trivial. This behavior originates in the fact that in order to define it we need the projectors Eq. 
\eqref{nnproj}. The definition of these projectors requires the null-vector $\tilde{n}$ that introduces another 
preferred direction (apart from the direction of $n$) which further breaks $SIM(1)$ symmetry.

However, we can change the projection $\tilde{\phi}_-$ in such a way that it has better transformation properties
with respect to $SIM(1)$ group. We introduce an operator
\begin{equation}\label{qdef}
 \hat{q} = \frac{\fsl{n}\fsl{\partial}}{2n\cdot\partial}D,
\end{equation}
which has only one nonzero component
\begin{equation}\label{qdefn}
 \hat{q}_- = D_- - \frac{\partial_{-+}}{\partial_{++}}D_+.
\end{equation}
The improved $SIM(1)$ superfield is defined as
\begin{align}
 \hat{\phi}_- &= (\hat{q}_-\Phi)\big\vert_{\theta_+=0}.
\end{align}
The new $SIM(1)$ transformation rule
\begin{align}
 \hat{\phi}'_-(x',\theta') &= e^{-A}\hat{\phi}_-(x,\theta),
\end{align}
does not suffer from mixing with the other superfield $\phi$. 

In the case of gauge theory we are going to replace the derivatives in Eq. \eqref{qdef} with covariant ones.  
The covariant derivatives in the Lorentz invariant theory are given by 
\begin{align}
 \nabla_\alpha &= D_\alpha -i \Gamma_\alpha,
&
 \nabla_{\alpha\beta} &= \partial_{\alpha\beta} -i \Gamma_{\alpha\beta},
\end{align}
such that the (anti)commutators are given by 
\begin{align}
 \{\nabla_\alpha,\nabla_\beta\} &= -2\nabla_{\alpha\beta},
\nonumber\\
 [\nabla_\alpha,\nabla_{\beta\gamma}] &= C_{\alpha(\beta} W_{\gamma)},
\nonumber\\
 [\nabla_{\alpha\beta},\nabla_{\gamma\delta}] &= -\frac{1}{2} C_{\alpha\gamma} F_{\beta\delta}
	-\frac{1}{2} C_{\alpha\delta} F_{\beta\gamma} 
	-\frac{1}{2} C_{\beta\delta} F_{\alpha\gamma}
	-\frac{1}{2} C_{\beta\gamma} F_{\alpha\delta},
\end{align}
where
\begin{align}
 \Gamma_{\alpha\beta} &= -\frac{1}{2}\left( D_{(\alpha}\Gamma_{\beta)} -i \{\Gamma_\alpha,\Gamma_\beta\} \right),
\nonumber\\
 W_\alpha &= -\frac{i}{2}D^\beta D_\alpha \Gamma_\beta - \frac{1}{2}[\Gamma^\beta,D_\beta\Gamma_\alpha]
  + \frac{i}{6}[\Gamma^\beta,\{\Gamma_\beta,\Gamma_\alpha\}],
&
 \nabla^\alpha W_{\alpha} &= 0,
\nonumber\\
 F_{\alpha\beta} &= \frac{1}{2}\nabla_{(\alpha} W_{\beta)}.
\end{align}
There is more than one way how to define covariant version of the operator Eq. \eqref{qdef} because 
covariant derivatives do not commute among each other so the definition of this operator 
is ordering dependent.
In this text we will use the following variant
\footnote{
In $3+1$ dimensions \cite{supg} the ordering ambiguity is resolved if we want the $\phi_-$ projection of a covariantly 
chiral superfield $\bar{\nabla}_{\dot{\alpha}}\Phi=0$ satisfy the $SIM(2)$ chiral-covariant condition 
$\bar{\nabla}_{\dot{+}}\phi_-=0$. This forces us to choose the ordering
\begin{equation}
 \phi_- = \left(\left(\nabla_- - \nabla_{-\dot{+}}\frac{1}{\nabla_{+\dot{+}}}\nabla_+\right)\Phi\right)
 \big\vert_{\theta_+=0,\bar{\theta}_{\dot{+}}=0}.\nonumber 
\end{equation}
}
\begin{equation}
 q = \fsl{n}\fsnabla\frac{1}{2n\cdot\nabla}\nabla,
\end{equation}
which leads to the definition of the superfield
\begin{equation}
 \phi_- = \left(q_-\Phi\right)\vert_{\theta_+=0} = \left(\left(\nabla_- - \nabla_{+-}\frac{\nabla_+}{\nabla_{++}}\right)\Phi\right)\big\vert_{\theta_+=0}.
\end{equation}
The $SIM(1)$ transformation properties are the same as in the case of $\hat{\phi}_-$.

In Eq. \eqref{qdefn},  we introduced the nonlocal operator $\frac{1}{\partial_{++}}$. A similar operator 
$\frac{1}{\partial_{+\dot{+}}}$ appears in the four-dimensional $SIM(2)$ theory, the properties of this operator were
discussed in detail in \cite{supg}, and the same arguments that were presented there apply also to our case. 
The operator $\frac{1}{\partial_{++}}$ has to be linear and satisfy the condition 
$\partial_{++}\tfrac{1}{\partial_{++}}=1$, i.e. it is a propagator associated with $\partial_{++}$.
In addition to that we require it to commute with space-time derivatives. This is a nontrivial requirement because
the condition that it commutes with $\partial_{++}$ gives
\begin{multline}\label{pcom}
 \left[\frac{1}{\partial_{++}},\partial_{++}\right]f(x)
 = \left(\frac{1}{\partial_{++}}\partial_{++} - \partial_{++}\frac{1}{\partial_{++}}\right)f(x)  
 = \frac{1}{\partial_{++}}\partial_{++} f(x) - f(x) = 0.
\end{multline}
But this is evidently not true for nonzero functions satisfying $\partial_{++} f(x)=0$. The solution to this problem is
to restrict the space of functions to those that satisfy the condition Eq. \eqref{pcom}. 
One way how to define this operator is (omitting anticommuting coordinates)
\begin{equation}
 \frac{1}{\partial_{++}}f(x^{++},x^{--},x^{+-}) = 
  \int_{-\infty}^{x^{++}}\di t^{++}f(t^{++},x^{--},x^{+-}).
\end{equation}
and restrict the space of functions to those that satisfy $\lim_{x^{++}\rightarrow -\infty}f(x^{++},x^{--},x^{+-})=0$.
One of consequences of the fact that we are working with the reduced space of functions is 
that the equation $\partial_{++} f(x)=0$ has only one solution $f(x)=0$.
The covariant version of the operator $\frac{1}{\nabla_{++}}$ should retain most of the properties of the operator 
$\frac{1}{\partial_{++}}$, it should be linear, inverse to $\nabla_{++}$ and commute with $\nabla_{++}$. 
We cannot require it to commute with other covariant derivatives because covariant derivatives do not commute among 
each other. As in general formalism the explicit expression for the covariant derivative is not given, 
we do not construct an explicit expression for this operator. However, an explicit expression for this operator is not 
needed for obtaining the main results of this paper.

If the $SIM(1)$ superfield carries space-time indices we arrive at basically the same problem as in the case of 
$q_-$. We will illustrate this problem on the superfield $W_\alpha$  
\footnote{
We should also consider $\tilde{q}_-$ projections, but in the case of field strengths in gauge theory we do not need 
them.
}. 
The projections
\begin{align}\label{wtproj}
 w_+ &= W_+\vert_{\theta_+ = 0},
&
 \tilde{w}_- &= W_-\vert_{\theta_+ = 0},  
\end{align}
transform under the action of $SIM(1)$ group as
\begin{align}\label{wtsimt}
 w'_+(x',\theta') &= e^{A} w_+(x,\theta),
&
 \tilde{w}'_-(x',\theta') &= e^{-A} \tilde{w}_-(x,\theta) + B w_+(x,\theta).
\end{align}
the projection $\tilde{w}_-$ has the same ugly transformation rule as we had for $\tilde{\phi}_-$.
The transformation properties can be improved by the same trick that we used above. We introduce an operator
\footnote{
The ordering in this definition was chosen in this way because it results in a very simple rule for integration by parts
\begin{equation}
 \int \di^3x \nabla_+ \left((\triangle_{\alpha\beta}f)g\right) = \int \di^3x \nabla_+ \left(f(\triangle_{\alpha\beta}g)\right)
 + \text{surface terms}.\nonumber 
\end{equation}
where $f$ and $g$ are arbitrary superfunctions.
}
\begin{align}
 \triangle = \frac{i}{2}\left(\frac{1}{2n\cdot\nabla}\fsl{n}\fsnabla + \fsl{n}\fsnabla\frac{1}{2n\cdot\nabla}\right).
\end{align}
the only nonzero components of $\triangle_{\alpha\beta}$ are
\begin{align}\label{deftriangle}
 \triangle_{-+} &= 1,
&
 \triangle_{--} &= \frac{1}{2}\nabla_{+-}\frac{1}{\nabla_{++}} +\frac{1}{2}\frac{1}{\nabla_{++}}\nabla_{+-}.
\end{align}
We define
\begin{equation}
 w_- = i(\triangle_-{}^\alpha W_\alpha)\vert_{\theta_+ = 0} = (W_- - \triangle_{--}W_+)\vert_{\theta_+ = 0}.
\end{equation}
If there are more space-time indices we have to repeat this procedure for each index, in particular we will need to do 
this for the superfield $F_{\alpha\beta}$. We define
\begin{align}
 f_{++} &= F_{++}\vert_{\theta_+ = 0},
\nonumber\\
 f_{+-} &= i(\triangle_-{}^\alpha F_{+\alpha})\vert_{\theta_+ = 0} = (F_{+-} - \triangle_{--}F_{++})\vert_{\theta_+ = 0},
\nonumber\\
 f_{--} &= -(\triangle_-{}^\alpha \triangle_-{}^{\beta} F_{\alpha\beta})\vert_{\theta_+ = 0} = (F_{--} - 2\triangle_{--}F_{+-} + \triangle_{--}\triangle_{--}F_{++})\vert_{\theta_+ = 0},
\end{align}
The superfields that we obtain in this way have very simple transformation properties under action of $SIM(1)$.
For a general superfield $\psi_{+\cdots+-\cdots-}$ we can schematically write this rule as
\begin{align}\label{tsimc}
 \psi'_{+\cdots+-\cdots-}(x',\theta') &= e^{A(\text{\# of ``+'' indices minus \# of ``-'' indices})}\psi_{+\cdots+-\cdots-}(x,\theta).
\end{align}

\section{Gauge theory with $\mathcal{N} = 1/2$ supersymmetry}
In this section, we will use $SIM(1)$ superspace to study super-Yang-Mills theory coupled to matter field. 
As the reduction of the $\mathcal{N} =1$ superspace to $SIM(1)$ superspace breaks the supersymmetry from 
$\mathcal{N} =1$ supersymmetry to $\mathcal{N} =1/2$ supersymmetry, the Yang-Mills theory coupled to matter fields
will have $\mathcal{N} =1/2$ supersymmetry. It may be noted that if we do not break the Lorentz symmetry but use the 
$SIM(1)$ formalism for analysing the super-Yang-Mills theory coupled to matter field, 
  we can recover the full    
$\mathcal{N}=1$ supersymmetry. In this case 
  only   $\mathcal{N}=1/2$ supersymmetry will be manifested in the  superspace formalism. The other half of 
the symmetry can be considered as an accidental symmetry that would disappear
once we use this formalism in its intended role -- to study effects
that break the Lorentz symmetry but preserve $SIM(1)$ symmetry.

We will consider the a Lorentz invariant action for matter superfields as  
\begin{equation}\label{slorm}
 S_m = \frac{1}{2}\int \di^3x \nabla^2 \left[ \big(\nabla^\alpha\Phi^\dagger\big) \big(\nabla_\alpha\Phi\big)  \right], 
\end{equation}
and a Lorentz invariant action for the gauge superfield as 
\begin{equation}
 S_g = \text{tr} \int \di^3x \nabla^2 \left[ W^2 \right].
\end{equation}
In this section we are going to write down these actions in $SIM(1)$ formalism. This would not make much sense
if the Lorentz symmetry was not broken because in that case the ordinary superspace would provide more convenient 
setting. However, these actions can also serve as a basis for theories where the Lorentz symmetry is broken and
the Lorentz invariant formalism does not provide adequate setting. Some of Lorentz breaking mechanisms are discussed
in the next sections.
When the space-time covariant derivatives appear in the $SIM(1)$ action, we have to understand them as 
projections $\nabla_{\alpha\beta}\vert_{\theta_+=0}$ of $SO(2,1)$ derivatives.
The matter field action in the $SIM(1)$ superspace formalism is
\begin{multline}\label{sacm}
 S_m = \frac{1}{2}\int\di^3x \nabla_+ \Bigg[
    \big(\nabla_+\phi^\dagger_-\big) \phi_-
  + \phi^\dagger_-\big (\nabla_+\phi_-\big)
  -2i \phi^\dagger_- \left(w_+\frac{\nabla_+}{\nabla_{++}}\phi\right)
  -2i \left(w_+\frac{\nabla_+}{\nabla_{++}}\phi^\dagger\right) \phi_-
  \\
  + \big(\Box_\text{cov}\phi^\dagger\big) \left(\frac{\nabla_+}{\nabla_{++}}\phi\right)
  + \left(\frac{\nabla_+}{\nabla_{++}}\phi^\dagger\right) \big(\Box_\text{cov}\phi\big)
  \\
  -i \left(\frac{\nabla_+}{\nabla_{++}}\phi^\dagger\right) \left(\big(\nabla_{++}w_-\big)\frac{\nabla_+}{\nabla_{++}}\phi\right)
  + \frac{1}{2} \left(\frac{\nabla_+}{\nabla_{++}}\phi^\dagger\right) \left(\left[f_{++},\frac{1}{\nabla_{++}}w_+\right]\frac{\nabla_+}{\nabla_{++}}\phi\right)
  \Bigg]
  \\+ \text{surface terms},
\end{multline}
and we can say that it is explicitly $SIM(1)$ invariant. 
In fact each term that appears in the action is separately $SIM(1)$ invariant.  
The covariant d'Alambertian operator is defined as $\Box_\text{cov} = -\frac{1}{2}\nabla^{\alpha\beta}\nabla_{\alpha\beta}$.

Now we can write the action for gauge sector of the theory as
\begin{equation}
 S_g = \text{tr} \int \di^3x \nabla_+ \left( -\tilde{f}_{+-}\tilde{w}_- + w_+\tilde{f}_{--} \right) + \text{surface terms}.
\end{equation}
This form, that uses projectors Eq. \eqref{wtproj}, does not show manifest $SIM(1)$ invariance.  
The $SIM(1)$ transformations change, according to Eq. \eqref{wtsimt}, this action to 
\begin{multline}
 S'_g = S_g + (e^AB)\text{tr} \int \di^3x \nabla_+ \left(w_+\tilde{f}_{+-} - \tilde{f}_{++}\tilde{w}_-\right)\\
 + (e^{2A}B^2)\text{tr} \int \di^3x \nabla_+ \left(w_+f_{++} - f_{++}w_+\right)
\\
 = S_g + (e^AB)\text{tr} \int \di^3x \nabla_+ \left(-\nabla_+\left(w_+\tilde{w}_{-}\right)\right) 
 = S_g + \text{surface terms},
\end{multline}
where we used $\nabla_+w_+ = f_{++}$ and $\nabla_+\tilde{w}_- = \tilde{f}_{+-}$. We see that the $SIM(1)$ invariance is not obvious at first glance.
This is a reason why it is better to write the action in terms of $SIM(1)$ superfields that have simple transformation properties
\begin{multline}\label{sacg}
 S_g = \text{tr} \int \di^3x \nabla_+ \bigg[ -f_{+-}w_- + w_+f_{--} 
 -\frac{i}{2}\left\{w_+,\frac{1}{\nabla_{++}}w_+\right\}w_- -\frac{i}{2}\left(\frac{1}{\nabla_{++}}\{w_+,w_+\}\right)w_-\\
 -\frac{1}{2}\left(\triangle_{-}{}^\alpha w_+\right)\left\{w_+,\frac{1}{\nabla_{++}}(\triangle_{-\alpha}w_+)\right\}
 \bigg] + \text{surface terms}.
\end{multline}
Each term in this action is separately $SIM(1)$ invariant. The verification of the $SIM(1)$ invariance is easy because the
superfields $w_+$, $w_-$, $f_{++}$, $f_{+-}$, $f_{--}$ and derivatives $\nabla_+$, $\nabla_{++}$ transform under 
$SIM(1)$ according to the rule Eq. \eqref{tsimc}. Thus the expressions are invariant if they contain the same number of 
lower plus indices as there are lower minus indices. The only exception to this rule are the operators 
$\triangle_{-\alpha}$ that appear in the last term. This term is equal to
\begin{equation}
 -\frac{i}{2}\left(\triangle_{--}w_+\right)\left\{w_+,\frac{1}{\nabla_{++}}w_+\right\} +\frac{i}{2}w_+\left\{w_+,\frac{1}{\nabla_{++}}(\triangle_{--}w_+)\right\}
\end{equation}
where we used that $\triangle_{-+}=1$. Using the transformation rule 
$\triangle_{--} \rightarrow e^{-2A}\triangle_{--}+e^{-A}B\triangle_{-+} = e^{-2A}\triangle_{--}+e^{-A}B$
we find that $SIM(1)$ transformations change this term as
\begin{multline}
 \delta\left(\text{tr} \int \di^3x \nabla_+ \left[
 -\frac{i}{2}\left(\triangle_{--}w_+\right)\left\{w_+,\frac{1}{\nabla_{++}}w_+\right\} +\frac{i}{2}w_+\left\{w_+,\frac{1}{\nabla_{++}}(\triangle_{--}w_+)\right\}
 \right]\right)
\\
 = e^A B \,
 \text{tr} \int \di^3x \nabla_+ \left[
 -\frac{i}{2}w_+\left\{w_+,\frac{1}{\nabla_{++}}w_+\right\} +\frac{i}{2}w_+\left\{w_+,\frac{1}{\nabla_{++}}w_+\right\}
 \right]
 = 0,
\end{multline}
so this term is also $SIM(1)$ invariant.
Thus,  we have been able to write the action of super-Yang-Mills theory coupled to matter fields in $SIM(1)$ superspace. 

The actions Eq. \eqref{sacm} and Eq. \eqref{sacg} contain nonlocal operator $\frac{1}{\nabla_{++}}$, 
but that does not mean that they describe nonlocal theory. In fact, we know that the actions 
Eq. \eqref{sacm} and Eq. \eqref{sacg} describe local theory because they are derived from local Lorentz invariant 
actions. 
A four-dimensional supersymmetric theory provide us with another example where we encounter nonlocal operators 
in a local theory. When we write a chiral integral in a form with integral over full superspace we obtain an expression 
that contains a nonlocal operator
\footnote{
For example, assume that $\Phi$ is a chiral superfield. The chiral integral of $\Phi^2$ could be written as
\begin{equation*}
 \int \di^4 x D^2 \left(\Phi^2\right) = \int \di^4 x D^2 \bar{D}^2 \left(\Phi\frac{D^2}{\Box}\Phi\right).
\end{equation*}
}.
In the same way operators $\frac{1}{\nabla_{++}}$ play a very similar role in $SIM(1)$ superspace.
This does not imply that any theory in $SIM(1)$ superspace is non-local, just as the existence 
of a non-local operator in the four-dimensional chiral superspace does not imply that any theory in chiral superspace 
is non-local. In absence of a Lorentz breaking term, we could still write the 
  action of a local three dimensional theory with 
$\mathcal{N}=1$ supersymmetry in $SIM(1)$ superspace, in which only half of supersymmetry is manifest. 
This is just a complicated way to write the original action with $\mathcal{N}=1$ supersymmetry. 
Now, as the original action was local, the same action written in $SIM(1)$ superspace has to also be local, 
despite the presence of nonlocal operators. 

So far we have worked with a particular choice of the vector $n$, but we could also write the results in a form that 
shows explicit dependence on this vector. 
Thus, the action for the matter sector can be written as 
\begin{multline}
 S_m = \frac{1}{2} \int \di^3x \nabla^\alpha \bigg[
 - \left(\nabla^\beta q_\beta\Phi^\dagger\right) \left(q_\alpha\Phi\right)
 - \left(q_\alpha\Phi^\dagger\right) \left(\nabla^\beta q_\beta\Phi\right)
 \\
 + 2\left(q_\alpha\Phi^\dagger\right) \left(W^\beta\frac{n_\beta{}^\gamma}{\sqrt{2}n\cdot\nabla}\nabla_\gamma\Phi\right)
 + 2 \left(W^\beta\frac{n_\beta{}^\gamma}{\sqrt{2}n\cdot\nabla}\nabla_\gamma\Phi^\dagger\right) \left(q_\alpha\Phi\right)
 \\
 + \left(\Box_\text{cov}\Phi^\dagger\right) \left(\frac{n_\alpha{}^\beta}{\sqrt{2}n\cdot\nabla}\nabla_\beta\Phi\right)
 + \left(\frac{n_\alpha{}^\beta}{\sqrt{2}n\cdot\nabla}\nabla_\beta\Phi^\dagger\right) \left(\Box_\text{cov}\Phi\right)
 \\
 -i \left(\frac{n_\alpha{}^\beta}{\sqrt{2}n\cdot\nabla}\nabla_\beta\Phi^\dagger\right) \left(\left((\sqrt{2}n\cdot\nabla)\triangle^{\gamma\delta}W_\delta\right)\frac{1}{\sqrt{2}n\cdot\nabla}\nabla_\gamma\Phi\right)
 \\
 +\frac{1}{2} \left(\frac{n_\alpha{}^\beta}{\sqrt{2}n\cdot\nabla}\nabla_\beta\Phi^\dagger\right) \left(\left[n^{\gamma\delta}f_{\gamma\delta},\frac{1}{\sqrt{2}n\cdot\nabla}W_\sigma\right]\frac{n^{\sigma\epsilon}}{\sqrt{2}n\cdot\nabla}\nabla_\epsilon\Phi\right)
 \bigg] + \text{surface terms},
\end{multline}
and the action for the gauge sector  can be written as 
\begin{multline}
 S_g = \text{tr} \int \di^3x \nabla^\alpha \bigg[
   \left(\triangle^{\gamma\delta} F_{\gamma\delta}\right)\left(\triangle_\alpha{}^\beta W_\beta\right)
 + \left(\triangle_\alpha{}^\gamma\triangle_\beta{}^\delta F_{\gamma\delta}\right) W^\beta
 \\
 - \frac{i}{2}\left\{W_\gamma,\frac{n^{\gamma\delta}}{\sqrt{2}n\cdot\nabla}W_\delta\right\}\left(\triangle_\alpha{}^\beta W_\beta\right)
 - \frac{i}{2}\left(\frac{n^{\gamma\delta}}{\sqrt{2}n\cdot\nabla}\left\{W_\gamma,W_\delta\right\}\right)\left(\triangle_\alpha{}^\beta W_\beta\right)
 \\
 - \frac{1}{2}\left(\triangle^{\gamma\beta}W_\gamma\right)\left\{W_\sigma,\frac{n_\alpha{}^\sigma}{\sqrt{2}n\cdot\nabla}\left(\triangle^\delta{}_\beta W_\delta\right)\right\}
 \bigg] + \text{surface terms}.
\end{multline}
The fact that we could write the action in this form proves that the supersymmetry is broken only due to presence of 
the preferred light-like direction determined by $n$. 

\section{Examples of Lorentz symmetry breaking}

This section is devoted to the discussion of two simple examples of Lorentz symmetry breaking. 
In each example the origin of Lorentz symmetry breaking will be different, in the first case it will be
a contribution to the action which violates Lorentz symmetry and in the second case it will be a presence of a boundary.

We may consider a Lorentz breaking contribution to the action that has a form
\begin{equation}\label{brlcg}
 S_b = \int \di^3x D_+ \mathcal{L}_-
 = \int \di^3x d_+ \left( \mathcal{L}_-\vert_{\theta_+=0} \right),
\end{equation}
where the Lorentz breaking Lagrangian $\mathcal{L}$ transforms under the $SIM(1)$ group as   
\begin{equation}
  \mathcal{L}'_-(x',\theta')=e^{-A} \mathcal{L}_-(x,\theta). 
\end{equation}
 This ensures invariance with respect  
to $SIM(1)$ rotations, invariance with respect to space-time translations is ensured by integral over space-time.
The only thing that remains to be checked is invariance with respect to supersymmetry transformations. 
The change caused by infinitesimal supersymmetry transformation is  
\begin{multline}
 \delta S_b = \int \di^3x D_+ \left( \delta \mathcal{L}_- \right)
 = \int \di^3x D_+ \left( -\epsilon^\alpha Q_\alpha \mathcal{L}_- \right)
 \\
 = \epsilon^\alpha \int \di^3x D_+ \left( (D_\alpha + 2\theta^\beta \partial_{\beta\alpha}) \mathcal{L}_- \right)
 = \epsilon^- \int \di^3x D_+ D_- \mathcal{L}_-,
\end{multline}
where $\epsilon^\alpha$ are infinitesimal anticommuting parameters.
In the last equality we used the fact that all surface terms vanish. We see that only supersymmetry transformations 
with $\fsl{n}\epsilon=\epsilon^-=0$ leave Eq. \eqref{brlcg} unchanged. This is the same condition that we used to break
the $\mathcal{N}=1$ supersymmetry to $\mathcal{N}=1/2$ supersymmetry.

An example of a Lorentz breaking contribution to the action that has this form is a Lorentz breaking mass term for 
superfield $\Phi$
\begin{equation}
 S_b = - m^2 \int \di^3x \nabla_+ \left( \phi^\dagger \frac{\nabla_+}{\nabla_{++}}\phi \right) 
 = m^2 \int \di^3x \nabla_\alpha\left( \Phi^\dagger \frac{n^{\alpha\beta}}{\sqrt{2}n\cdot\nabla} \nabla_\beta \Phi\right).
\end{equation}
In component form we get
\begin{multline}
 S_m + S_b = \int \di^3x \bigg(
 - A^\dagger (\Box_{\text{cov}}-m^2)A
 - \psi^{\dagger\alpha} \left(\nabla_{\alpha\beta} - m^2\frac{n_{\alpha\beta}}{\sqrt{2}n\cdot\nabla} \right) \psi^\beta
 \\
 - A^\dagger W^\alpha \psi_\alpha + \psi^{\dagger\alpha} W_\alpha A
 - F^\dagger F
 \bigg),
\end{multline}
where $A=\Phi\vert_{\theta=0}$, $\psi_\alpha = (\nabla_\alpha\Phi)\vert_{\theta=0}$ and 
$F=(\nabla^2\Phi)\vert_{\theta=0}$ are projections of $\Phi$.
It may be noted that when we introduce a Lorentz  breaking contribution we can write the action
only as a total $\nabla_+$ derivative. It is not possible to write this action as a total $\nabla^2$ derivative. 
On the other hand, it is possible to write a Lorentz invariant theory in $SIM(1)$ superspace, and in this case, half the 
supersymmetry of the theory will remain hidden.
However, it is not possible to express a theory with  $SIM(1)$ symmetry in the original superspace.

Now, we are going to look at another mechanism of Lorentz symmetry breaking.
We are going to consider a boundary consisting of points that satisfy the condition $n\cdot x = 0$, which in our choice 
of $n$ means that $x^{--}=0$. The space-time symmetry of such set of points consist of $SIM(1)$ rotations 
and translations generated by $P_{+-}$, $P_{--}$. The symmetry generator $P_{--}$ does not generate transformation 
preserving the boundary. Thus, the space-time symmetry that we use in this case is little different from what we 
considered in section \ref{susy3}. The boundary condition that we are going to use is that the superfield $\Phi$
vanishes for $n\cdot x = 0$
\begin{equation}
 \Phi \vert_{x^{--}=0} = 0.
\end{equation}
While the space-time symmetry was determined by the shape of the boundary surface, the amount of unbroken supersymmetry 
will follow from the requirement that the boundary condition is invariant. 
The infinitesimal supersymmetry transformation change the boundary condition as
\begin{multline}
 \delta \Phi \vert_{x^{--}=0} = -(\epsilon^\alpha Q_\alpha \Phi) \vert_{x^{--}=0}
 \\
 = -\left[ 
  \epsilon^+(\partial_+ + \theta^+ \partial_{++} + \theta^- \partial_{+-}) \Phi
  + \epsilon^-(\partial_- + \theta^+ \partial_{+-} + \theta^- \partial_{--}) \Phi
  \right] \vert_{x^{--}=0}
 \\
 = -\epsilon^- ( \theta^- \partial_{--} \Phi ) \vert_{x^{--}=0}.
\end{multline}
Thus, we are again forced to limit supersymmetry transformations to those that satisfy $\fsl{n}\epsilon=\epsilon^-=0$.

In both of our examples it was not enough to break space-time symmetry to $SIM(1)$, we also had to break half of 
supersymmetry.

\section{Conclusion}
In this paper, we analysed   three dimensional super-Yang-Mills theory in $SIM(1)$ superspace. 
 The original  Lorentz invariant theory  had  $\mathcal{N} =1$ supersymmetry. 
  However, when the Lorentz symmetry was broken down to $SIM(1)$ group,  the resultant 
theory preserved only half the supersymmetry of the original theory. As the original theory had  $\mathcal{N} =1 $, 
   so, the theory in $SIM(1)$ superspace has  $\mathcal{N} =1/2$ supersymmetry. This was the first time 
that   $\mathcal{N} =1 $ supersymmetry was broken down to $\mathcal{N} =1/2$ supersymmetry in 
three dimensions, on a manifold without a boundary. This is because for a manifold without a boundary, the  
other way to obtain a theory with $\mathcal{N} =1/2$ supersymmetry
is by imposing non-anticommutativity. However, in three dimensions there are not enough superspace degrees to allow this partial breaking 
of supersymmetry. So, any non-anticommutative deformation of a three dimensional theory with $\mathcal{N} =1 $ supersymmetry, 
will break all the supersymmetry of the resultant theory.  
It would be interesting to analyse a theory on a manifold with boundaries with  $SIM(1)$ superspace. 
It is known that the presence of a  boundary also breaks half the supersymmetry of a theory \cite{b2}-\cite{bou}. 
It is possible that both the boundary effects and 
the modification of the superspace to $SIM(1)$ superspace will break the same supercharges and hence will preserve half the supersymmetry of the 
original theory. It is also possible that a similar effect can be generated by studying non-anticommutativity in $SIM(1)$ superspace. 
It may be noted  that the   Wess-Zumino model with a  Lorentz symmetry breaking  term  has been quantized
in $SIM(2)$ superspace, and the one-loop effective action for this theory has also been constructed   \cite{supa}.
So, it would be interesting to analyse the quantization of three dimensional gauge theories in $SIM(1)$ superspace.

Three dimensional superspace is important as it has been used for studding three dimensional 
superconformal field theories.
Three dimensional superconformal field theory with $\mathcal{N} =8$ supersymmetry is thought to describe the low energy action 
for multiple $M2$-branes. This is because  apart from a constant closed 7-form on $S^7$,
 $AdS_4 \times S_7 \sim SO(2,3)\times SO(1,2)/ SO(8)\times SO(7) \subset OSp(8|4)/SO(1,3) \times SO(7)$, 
 and so, 
 $OSp(8|4)$ symmetry of the eleven dimensional supergravity on $AdS_4 \times S_7$ gets realized 
as $\mathcal{N} = 8$ supersymmetry of its dual superconformal field theory.
There are further constraints on this superconformal field theory which are satisfied by a 
theory called the BLG theory \cite{1}-\cite{aaaaa}. However, the gauge symmetry of the BLG theory is generated by 
a Lie $3$-algebra,  and   it only describes two M2-branes. 
It is possible to generalize the BLG theory to a theory describing any number of M2-branes and this theory is called the ABJM theory 
\cite{aaaa}-\cite{ab1}.  
Even though the ABJM theory has only $\mathcal{N} =6$ supersymmetry, it is expected that its supersymmetry might get enhanced 
to full $\mathcal{N} =8$ supersymmetry  \cite{abjm2}-\cite{ab00}.  It is also possible to use a Mukhi-Papageorgakis novel higgs mechanism  
to obtain a theory of multiple D2-branes from a theory of multiple M2-branes \cite{mu}-\cite{mu1}. The gauge sector for the low energy action  
of multiple D2-branes is described 
by a super-Yang-Mills theory. As it is known that certain unstable string theory vacuum states break Lorentz symmetry \cite{1a}-\cite{2a},  
it will be interesting to analyse the action of multiple D2-branes in $SIM(1)$ superspace. It would also be interesting to analyse 
the theory of multiple M2-branes and the Mukhi-Papageorgakis novel higgs mechanism in $SIM(1)$ superspace.

\section*{Appendix}
A spinor $\theta_\alpha$ is real (Majorana), spinor metric is antisymmetric and imaginary, 
the rules for raising and lowering of spinor indices are
\begin{align}
 \theta^\alpha &= \theta_\beta C^{\beta\alpha},
&
 \theta_\alpha &= \theta^\beta C_{\beta\alpha}.
\end{align}
Gamma matrices are real (Majorana)
\begin{align}
 \{\gamma^a,\gamma^b\} &= 2\eta^{ab},
&
 (\gamma^a)^* &= \gamma^a,
\end{align}
with space-time metric $\eta$ having signature $-1,+1,+1$.
The notation with spinor indices is related to the notation with matrix multiplication by identifying
\begin{align}
 \theta &\sim \theta_\alpha,
&
 \bar{\theta} = \theta^\dagger i\gamma^0 = \theta^T C &\sim \theta^\alpha,
&
 C &\sim C^{\alpha\beta},
&
 C^{-1} &\sim C_{\alpha\beta},
&
 \gamma^a &\sim (\gamma^a)_\alpha{}^\beta.
\end{align}
We also define
\begin{align}
 \theta^2 &= \frac{1}{2}\bar{\theta}\theta = \frac{1}{2} \theta^\alpha \theta_\alpha,
&
 \fsl{v} &= v_a \gamma^a,
\nonumber\\
 \gamma^a_{\alpha\beta} &= (\gamma^a C^{-1})_{\alpha\beta} = (\gamma^a)_\alpha{}^\gamma C_{\gamma\beta},
&
 \gamma_a^{\alpha\beta} &= -(C \gamma_a)_{\alpha\beta} = (\gamma_a)_\gamma{}^\beta C^{\gamma\alpha}.
\end{align}
There is a lot of useful relations
\begin{align}
 (\theta_\alpha)^* &= \theta_\alpha,
&
 (\theta^\alpha)^* &= -\theta^\alpha,
\nonumber\\
 C_{\alpha\gamma} C^{\gamma\beta} &= \delta_\alpha^\beta,
&
 C_{\alpha\beta} &= -C_{\beta\alpha} = -C^*_{\alpha\beta},
\nonumber\\
 \partial_\alpha \theta^\beta &= \delta^\beta_\alpha,
&
 \theta_\alpha \theta_\beta &= -C_{\alpha\beta}\theta^2,
\nonumber\\
 \gamma^a_{\alpha\beta} &= \gamma^a_{\beta\alpha} = -(\gamma^{a}_{\alpha\beta})^*,
&
 \gamma_a^{\alpha\beta} &= \gamma_a^{\beta\alpha} = -(\gamma_a^{\alpha\beta})^*,
\nonumber\\
 \gamma^a_{\alpha\beta} \gamma_b^{\alpha\beta} &= -2\delta^a_b,
&
 \gamma^a_{\alpha\beta} \gamma_a^{\gamma\delta} &= -\delta^\gamma_{(\alpha}\delta^\delta_{\beta)}.
\end{align}
The explicit form of spinor metric and gamma matrices can be, for example, chosen as 
\begin{align}
 C^{\alpha\beta} &= \sigma_2 = C_{\alpha\beta},
&
 (\gamma^a)_\alpha{}^\beta &= (i\sigma_2,\sigma_1,-\sigma_3),
&
 \gamma_a^{\alpha\beta} &= \gamma^a_{\alpha\beta} = (i\mathbf{1},i\sigma_3,i\sigma_1).
\end{align}
The correspondence between spinor and vector indices for coordinates, derivatives and other vectors (represented by $n$)
\begin{align}
 x^{\alpha\beta} &= \frac{1}{2} \gamma_{a}^{\alpha\beta} x^a,
&
 \partial_{\alpha\beta} &= \gamma^{a}_{\alpha\beta} \partial_a,
&
 n^{\alpha\beta} &= \frac{1}{\sqrt{2}} \gamma_{a}^{\alpha\beta} n^a
\end{align}
or if we need the inverse relations
\begin{align}
 x^a &= -\gamma^a_{\alpha\beta} x^{\alpha\beta},
&
 \partial_a &= -\frac{1}{2} \gamma_a^{\alpha\beta} \partial_{\alpha\beta},
&
 n^a &= -\frac{1}{\sqrt{2}} \gamma^a_{\alpha\beta} n^{\alpha\beta}.
\end{align}
With these rules we have
\begin{align}
 \partial_{\alpha\beta} x^{\gamma\delta} &= -\frac{1}{2} \delta^\gamma_{(\alpha} \delta^\delta_{\beta)},
&
 \partial^{\alpha\gamma}\partial_{\beta\gamma} &= -\delta^\alpha_\beta \Box,
\nonumber\\
 D_\alpha\theta^\beta &= \delta_\alpha^\beta,
&
 D^2 \theta^2 &= -1.
\end{align}

\end{document}